\documentclass[mathleft
]{an}

\usepackage{graphicx}
\usepackage{times}
\usepackage{natbib}
\bibpunct{(}{)}{;}{a}{}{,}

\def\HD{HD\,8801}
\def\vsin{\makebox{{\em v }{\rm sin} {\em i}}}
\def\logg{log\textit{g}}
\def\teff{T$_{eff}$}
\def\vmic{v$_{mic}$}
\def\vrad{v$_{rad}$}
\def\gd{$\gamma$\,Dor}
\def\ds{$\delta$\,Sct}

\def\kms{kms$^{-1}$}
\def\cms2{cms$^{-2}$}
\begin{document}

% The following seven commands are intended for editorial usage and should be ignored by
% the author(s).
\Pagespan{789}{}% Document's page range.
% If second parameter is left empty, the last page is computed automatically.
\Yearpublication{2006}%
\Yearsubmission{2005}%
\Month{11}%
\Volume{999}%
\Issue{88}%
% \DOI{This.is/not.aDOI}%

%
   \title{Abundance analysis of the $\gamma$ Doradus - $\delta$ Scuti hybrid metallic line (Am) star HD\,8801\thanks{The spectra were obtained at the McDonald Observatory, University of Texas, USA}}

   \author{R. Neuteufel \inst{1}
          \and
          W. Weiss\inst{1}
          \and
          G. Handler \inst{2}
          }

   \institute{$^1$ Institute for Astronomy (IfA), University of Vienna,
              T\"urkenschanzstrasse 17, A-1180 Vienna\\ \email{richard.neuteufel@univie.ac.at, werner.weiss@univie.ac.at}\\
             $^2$ Copernicus Astronomical Center, ul. Bartycka 18, 00-716 Warsaw, Poland\\ \email{gerald@gatekeeper2.camk.edu.pl} }

   \date{submitted: April 2013}

  \abstract{Low frequency oscillation, typical for $\gamma$\,Doradus $g-$mode type stellar core sensitive pulsation, as well as higher frequency $\delta$\,Scuti type pulsation typical for $p-$modes, sensitive to the envelope, make \HD\ a remarkable hybrid pulsator with the potential to probe a stellar structure over a wide range of radius. In addition \HD\ is a rare pulsating metallic line (Am) star. We determined the astrophysical fundamental parameters to locate \HD\ in the H--R diagram. We analyzed the element abundances, paying close attention to the errors involved, and confirm the nature of \HD\ as a metallic line (Am) star. We also determined an upper limit on the magnetic field strength. Our abundance analysis is based on classical techniques, but uses for the final step a model atmosphere calculated with the abundances determined by us. We also discuss spectropolarimetric observations obtained for \HD. This object is remarkable in several respects. It is a non-magnetic metallic line (Am) star, pulsating simultaneously in $p$- and $g$-modes, but also shows oscillations with periods in between these two domains, whose excitation requires explanation. Overall, the pulsational incidence in unevolved classical Am stars is believed to be quite low; \HD\ does not conform to this picture. Finally, about 75\% of Am stars are located in short-period binaries, but there is no evidence that \HD\ has a companion.}


   \maketitle
%
%________________________________________________________________

\section{Introduction}

\begin{figure}
    \centering
    \includegraphics[width=7.5cm]{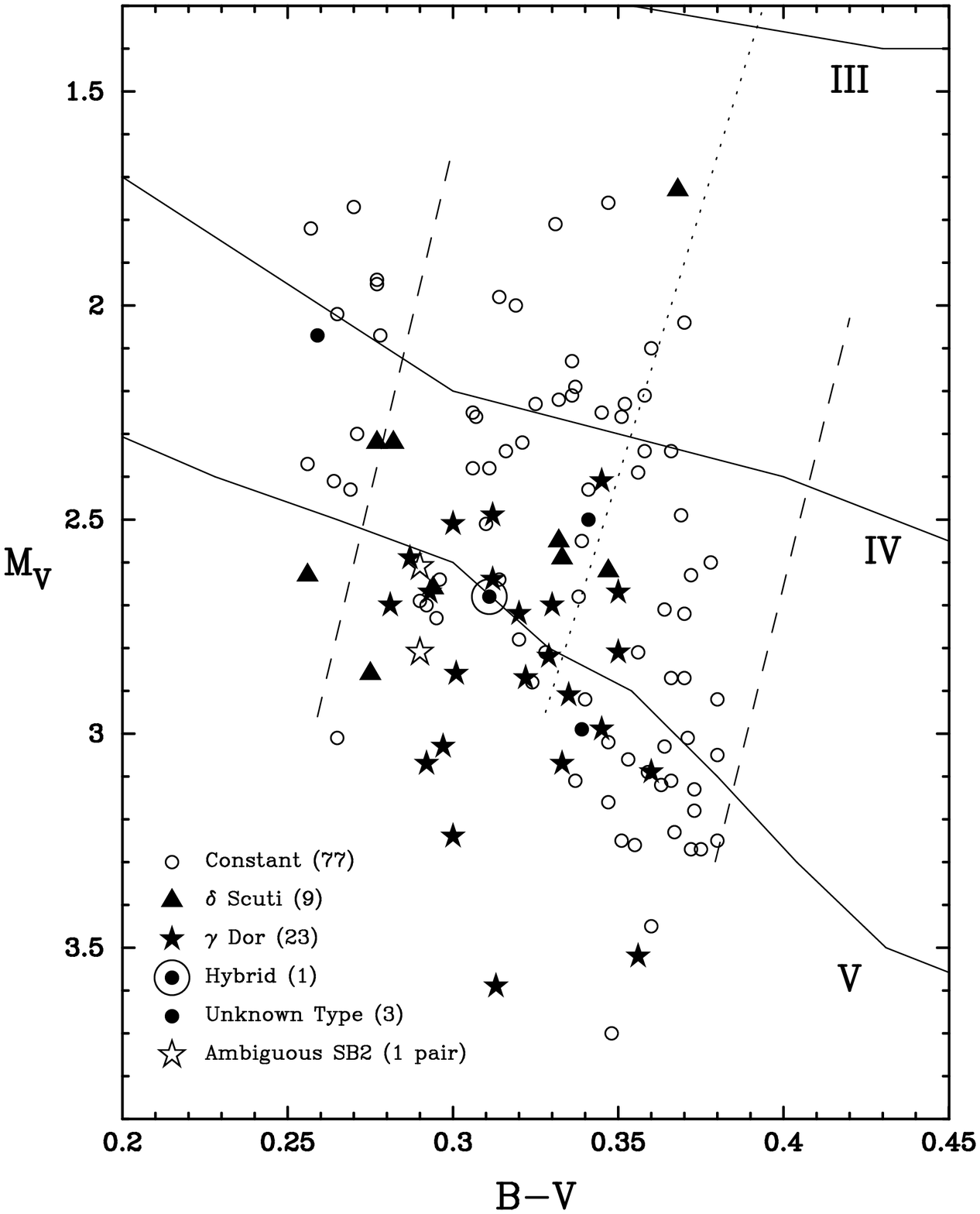}
          \caption{Position of \HD\ (\textit{large circled}) in the H-R diagram within both the \ds\ and \gd\ instability strips. This figure is adapted from \citet{2011AJ....142...39H}, and shows the 77 apparently constant and 37 variable stars examined during their survey. The luminosity classes, the \gd\ (dashed lines) and the red edge of the \ds\ instability strips (dotted line) also stem from \citet{2011AJ....142...39H}. Constant stars are plotted as open circles, \gd\ stars as filled stars, \ds\ stars as filled triangles and three variables of unknown type as filled circles.}
          \label{hrd8801}
    \end{figure}

    \HD is a main sequence star with temperature and luminosity in the overlapping region of the $\gamma$\,Doradus - $\delta$\,Scuti instability strips. Figure\,\ref{hrd8801} shows the \makebox{H-R} diagram reproduced by permission of the AAS taken from \citet{2011AJ....142...39H} with the position of \HD\ among other \gd\ stars.

    A SIMBAD database query resulted in the following values in the literature for Johnson colours V, (B-V) and (U-B) of 6.46, 0.27 and 0.03\,mag, Str\"{o}mgren colours (b-y), m1, c1 and H$\beta$ of 0.187, 0.196. 0.684 and 2.748\,mag, $\pi\,=\,17.91$\,mas, \vsin\,=\,55\,to\,79\,\kms, and spectral types ranging from A7m to F0m, which provided input for Table \ref{photometryparams}.

Low frequency oscillation were found in \HD, typical for \gd\ $g-$mode type pulsation, as well as higher frequency \ds\ type pulsation typical for $p-$modes, which make this star a remarkable hybrid pulsator \citep{2005AJ....129.2026H}, \citep{2009MNRAS.398.1339H}. With $g-$modes being sensitive to the stellar core and $p-$modes sensitive to the envelope, such hybrids potentially provide a unique possibility to probe a stellar structure over a wide range of radius as is argued, e.g. by \citet{2012A&A...540A.117C}.

Above all, \HD\ is a metallic line (Am) star although the incidence of pulsation in Am stars is lower than in chemically normal A/F stars. \citet{2000A&A...360..603T} show that Am stars become unstable only during evolution towards the red edge of the \ds\ instability strip, while diffusion processes in upper main sequence stars make young Am stars stable against pulsation \citep{1989MNRAS.238.1077K}, \citep{1998ASPC..135..420K}.

\citet{2011ApJ...743..153H} provided abundance analyses of two hybrid pulsators, HD 114839 and BD+18 4914, and showed the latter object to be an Am star. A spectroscopic signature of hybrid pulsation has not yet been found. \citet{2011ApJ...743..153H} therefore called for further abundance analyses of such pulsators, which we herewith provide.

    \begin{table}[h]
 \caption{Atmospheric parameters derived from Str\"{o}mgren and Johnson photometry extracted with SIMBAD and
 based on different calibrations using the software-tool \textit{TempLogG} \citep{2006ASPC..349..257K}.}
    \label{photometryparams}
    \centering
    \begin{tabular}{c c c c c}
    \hline\hline
    Str\"{o}mgren & M$_{v}$ & \teff & \logg & [Fe/H] \\
    photometry  & [mag] & [K] & & \\
    \hline
    (1) & 2.73 & 7250 & 4.13 & \\
    (2) & 2.73 & 6900 & 3.90 & \\
    (3) & 2.73 & 7370 & 4.09 & \\
    (4) & 2.73 & 7270 & 4.16 & \\
    (5) & 2.73 & 6700 & 3.46 & \\
    (6) & & & & 0.149 \\
    \hline\hline \\
     Johnson  & R& \teff & \logg & SpType \\
    photometry &[R/R$_{sun}$]  &  [K] & & \\
   \hline
    (7) & 1.497 & 7090 & 4.21 & F1 \\
    \end{tabular}
    \tablebib{Calibration from: (1) \citet{1985MNRAS.217..305M};
    (2) \citet{1993A&A...268..653N}; (3) \citet{1994MNRAS.268..119B}; (4) \citet{1997A&A...327..207R};
    (5) \citet{1997A&A...318..841C}; (6) \citet{1984A&AS...57..443O} and \citet{1989A&A...221...65S}; (7) \citet{1953ApJ...117..313J}.
    }
    \end{table}

After having prepared our data (Sec.\,\ref{preparing}), we start the abundance analysis with stellar fundamental parameter estimates based on photometry (\teff, \logg, and assumed \vmic\ and Z, see Table \ref{photometryparams}), improve iteratively \teff, \logg, and \vmic\ using mainly iron lines and model atmospheres based on ODF tables. With these parameters we determine to first order the abundance spectrum of \HD\ (Sec.\,\ref{modelparam}). In a next step we produce model atmospheres using opacities computed for particular abundance spectra (LL-models, see Sec.\,\ref{Tools}), determine the best fitting LL-model atmosphere and redo the abundance determination (Sec.\,\ref{abund}).

Finally we investigate the detectability of a magnetic surface field (Sec.\,\ref{magfield}), discuss the abundance errors of our analysis (Sec.\,\ref{error})  and present our results in section\,\ref{Results}.

%__________________________________________________________________

\section{Preparing for the abundance analysis}      \label{preparing}
    In this section we list and comment on the tools and steps needed to prepare the observations for an abundance analysis and to perform the analysis itself.

\subsection{Tools}\label{Tools}

\hspace{0mm}--\ \textit{TempLogG}
	dates back to E.M. Fresno \citep{1994} and estimates stellar fundamental parameters using published calibrations. The software was written in 2002 by Ch. St\"{u}tz and J. Nendwich, and was developed further by A. Kaiser \citep{2006ASPC..349..257K}. \\
\hspace{3mm}--\  \textit{Vienna Atomic Line Database (VALD):}
    Most important for stellar atmosphere analyses is the knowledge of accurate atomic parameters for as many spectral lines as possible. For our analysis we used VALD, which currently contains information on several million transitions compiled from dozens of individual catalogues (\citet{1995A&AS..112..525P}, \citet{1997BaltA...6..244R}, \citet{1999A&AS..138..119K}, \citet{2000BaltA...9..590K}, \citet{2008JPhCS.130a2011H}). VALD also provides tools for extracting spectral lines according to user specified criteria (\textit{Preselect}, \textit{Select}). \\
\hspace{3mm}--\ \textit{LLmodels:}
    The stellar atmosphere models used in our analysis were mostly calculated with the LL model code \citep{2004A&A...428..993S} which is used to compute atmospheres based on voluminous line lists representing individual abundance spectra. This mode was chosen for the final abundance determination, because it allows us to include the influence of the non solar abundance pattern on stellar atmospheres. However, for the first iterations we used the code in the ODF (opacity distribution function) mode which means that Kurucz ODF-tables based on solar scaled abundances were used for opacity calculations. This option increases the calculation speed by a factor of 100 compared to the LL (line-by-line) mode \citep{2005MSAIS...8..165S}.
    The full local spectrum turbulence convection model of \citet{1991ApJ...370..295C} is implemented in the code and was used throughout this analysis. \\
\hspace{3mm}--\ \textit{Synth3}
    is a fast spectrum synthesis code based on SYNTH written by N. Piskunov \citep{1992pess.conf...92P}. It calculates spectra based on a stellar model as produced e.g. by the LL-model code for static, plane parallel atmospheres in LTE \citep{2003ASPC..298..173N}. \\
\hspace{3mm}--\ \textit{Synth\_mag}
    is a synthesis code written by O. Kochukov \citep{2007pms..conf..109K} to synthesize spectra for a stellar atmosphere that includes a magnetic field, which is configured by radial, meridional and tangential components.
    The output is a spectrum for each of the four Stokes parameters. \\
\hspace{3mm}--\ \textit{Rotate\_OLEG}
    was developed by O. Kochukov and also produces a simultaneous graphic presentation of observed and synthetic spectra. The synthesis can be shifted in wavelength (radial velocity) and folded with a rotational broadening function until a best $\chi^2$-fit is found. \\
\hspace{3mm}--\ \textit{Atmospheric Tools Compilation (ATC)}
    is a software suite for computing stellar model atmospheres (LLmodels), spectrum synthesis (Synth3, Synth\_mag), and line-core-fitting (linfit). It was compiled by Ch. St\"{u}tz who also developed a user friendly interface.

\subsection{Observations}
    Two spectra were obtained by one of us (GH) with the 2.7\,m telescope at the McDonald Observatory on July 10, 2004, each with an integration time of 600 seconds. A cross-dispersed Echelle spectrometer was used at the Coud\'{e}-focus (f/32.5) with a resolution of 60$\,$000. A total of 62 Echelle orders covers the wavelength range from 3630\,{\AA} to 10275\,{\AA}, of which the orders close to the red and blue limits could not be used due to a very low S/N ratio, leaving the range from 4143\,{\AA} to 9193\,{\AA} available for our analysis except for the several {\AA}ngstr\"{o}ms wide gaps between the individual Echelle orders, which are typical for this instrument.

\subsection{Continuum normalization}
    The subtraction of  dark frames,  flatfielding,  order extraction and  wavelength calibration were done with standard IRAF routines. The two spectra were co-added and an envelope fitted as a first-order continuum presentation.

\begin{figure}
    \centering
    \includegraphics[width=7.5cm]{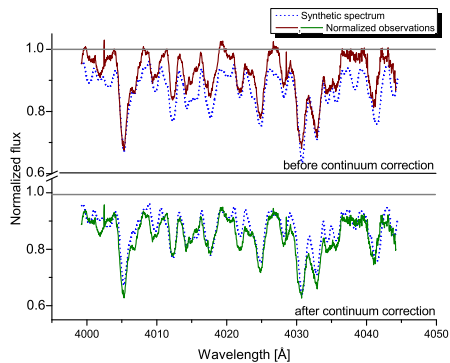}
          \caption{Comparison between a synthetic spectrum and a normalized observed spectrum before (upper part) and after the application (lower part) of the continuum correction procedure.}
          \label{renorm_beforeAfter}
    \end{figure}

    To improve the semi-automatic normalization the following procedure was developed and tested with synthetic spectra.
    All data points between 0.99 and 1.01 of relative intensity were flagged and the observed intensities compared to the corresponding intensities of a synthetic spectrum based on a model atmosphere with parameters estimated with the \citet{1985MNRAS.217..305M} calibration from Str\"{o}mgren photometry: \teff\,=\,7250\,K, \logg\,=\,4.13.
A polynomial up to order three was fitted through these flux differences, for each spectral order, defining a so-called ``continuum correction", which was then applied to the observation. Figure\,\ref{renorm_beforeAfter} shows a comparison between a synthetic spectrum and a small part of the original and finally corrected observations.

    Unfortunately, the Hydrogen lines could not be used as temperature indicators, because the respective Echelle orders could not be reliably continuum normalized.

\subsection{Radial velocity}\label{Radial velocity}

    \HD\ has a projected rotational velocity of about 55\,\kms, which adds to line blending and consequently impedes the radial velocity determination.  Throughout the whole spectrum with several thousands of contributing spectral lines, a visual inspection indicated only 40 usable, sufficiently blend-free lines. They are listed in Tab.\,\ref{vrad_individual} and give a mean radial velocity of $-22.7 \pm 1.7$\,\kms.

    In another approach we cross-correlated the continuum normalized observed Echelle orders with a synthetic spectrum based on an ODF-model, computed with the atmospheric parameters derived from the \citet{1985MNRAS.217..305M} calibration (see Tab.\,\ref{photometryparams}). The cross-correlations for all 26 Echelle orders were calculated in a RV interval from $-32.3$ to $-12.3$\,\kms\ with a step-width of 0.1\,\kms\ and resulted in  RV-values for the individual orders and a corresponding maximum correlation coefficient \textit{R$_{max}$}. Most of the orders contained high quality lines from Table~\ref{vrad_individual} (last column), with a maximum of four lines in a few orders.

We determined the average radial velocity of \HD\ first as an unweighted mean of all RV values obtained for the various Echelle orders (RV = $-23.2$\,\kms), a second time using \textit{R$_{max}$} as weight (RV = $-23.1$\,\kms), and finally using the number of high quality lines as a weight (RV = $-23.1$\,\kms). In the appendix we list he RV values for the individual orders in Tab.\,\ref{vrad} and the various mean values in Tab.\,\ref{vrad_ave}.

    Finally, a radial velocity of $-23.1$\,\kms was adopted for \HD. The corresponding heliocentric radial velocity is 1.4\,\kms, and the velocity relative to the LSR is 3.3\,\kms. This compares favourably with the value reported by \citep{2005AJ....129.2026H} from the same spectra, but using a smaller wavelength range, and the other radial velocities listed for HD~8801 by these authors.

    \begin{table}
    \caption{Results for \vrad\ and {\em v }{\rm sin} {\em i} by fitting observed to synthetic lines (tool \textit{Rotate\_OLEG}). oNr corresponds to the echelle order from Tab.\,\ref{vrad}. Units for \vrad\ and {\em v }{\rm sin} {\em i} are \kms. Multiple wavelengths indicate blended lines. In such a case, the whole spectral feature was fitted.}
    \label{vrad_individual}
    \centering
    \begin{tabular}{l c c c c}
    \hline\hline
    Element & $\lambda$ & \vrad & \vsin & oNr\\
    & [{\AA}] & & & \\
    \hline
    Fe I & 4005.24 & -25.6 & & 1 \\
    Fe I & 4063.59 & -24.2 & & 2 \\
    Ti II & 4163.64 & -22.9 & & 3 \\
    Fe I & 4191.43 & -25.0 & & 3 \\
    Fe I & 4222.21 & -20.2 & 50.4 & 4 \\
    Fe I & 4271.15, 4271.76 & -22.0 & 48.1 & 5 \\
    Fe I, Ti II & 4367.58, 4367.65 & -21.3 & & 6 \\
    Fe I & 4427.30, 4427.31 & -21.7 & 57.7 & 7 \\
    Ti II & 4501.27 & -23.9 & & 8\\
    Fe II & 4508.29 & -23.7 & 50.8 & 8 \\
    Fe II & 4515.34 & -21.0 & 49.9 & 8 \\
    Fe II & 4520.22 & -24.9 & 49.7 & 8 \\
    Fe II, Ti II & 4549.47, 4549.62 & -21.5 & 55.5 & 9 \\
    Cr II & 4558.65 & -19.9 & 49.3 & 9 \\
    Ti II & 4563.76 & -21.4 & 54.7 & 9 \\
    Ti II & 4571.97 & -18.6 & 48.2 & 9 \\
    Fe I & 4602.94 & -24.0 &  & 10 \\
    Fe I & 4957.30, 4957.60 & -22.9 & 54.3 & 11 \\
    Fe I & 5090.77 & -21.8 & 54.8 & 12 \\
    Fe I & 5162.27 & -21.4 & 51.4 & 12 \\
    Fe I & 5191.46, 4192.34 & -21.0 &  & 12 \\
    Ti II, Fe I & 5226.54, 5226.86 & -24.1 &  & 13 \\
    Fe I, Ca I,  & 5269.54, 5270.27,  &  &  & \\
    ... Fe I & 5270.36 & -25.2 & & 13 \\
    Fe I & 5302.30 & -24.9 & 54.4 & 14 \\
    Fe II & 5316.62, 5316.78 & -23.1 & 52.4 & 14 \\
    Fe I & 5353.37 & -22.5 & 53.3 & 14 \\
    Fe I & 5383.37 & -19.7 & 52.6 & 15 \\
    Fe I & 5400.50 & -23.3 & 54.4 & 15 \\
    Fe I & 5429.70 & -21.6 & 56.1 & 15 \\
    Fe I & 5497.52 & -24.9 & 53.4 & 16 \\
    Fe II & 5534.85 & -21.0 & 49.8 & 16 \\
    Fe I & 5569.62 & -22.6 & 52.5 & 17 \\
    Fe I & 5572.84 & -24.7 & 52.3 & 17 \\
    Fe I & 6065.48 & -23.2 & 54.4 & 20 \\
    Ca I & 6102.72 & -25.2 & 57.7 & 20 \\
    Fe I & 6191.56 & -22.8 & 53.0 & 21 \\
    Fe I & 6393.60 & -21.9 & 56.5 & 22 \\
    Fe I & 6411.65 & -22.3 & 54.1 & 22 \\
    Ca I & 6439.08 & -22.4 & 55.6 & 22 \\
    Fe I & 7495.07 & -24.0 & 56.0 & 25 \\
    \hline
    \end{tabular}
    \end{table}

\subsection{Projected rotational velocity}

    The tool \textit{Rotate\_OLEG} also provides a fitting algorithm to spectral line profiles for determining the projected rotational velocity. A synthetic spectrum with no rotational broadening is folded with a rotation profile and compared with observations until $\chi^2$ is minimized. 30 out of the 40 lines of Tab.\,\ref{vrad_individual} were suited for this analysis (see column \vsin\ in this Table), which resulted in a projected rotational velocity of 53.1\,$\pm$\,0.5\,\kms. For the remaining 10 lines either the continuum was obviously incorrect or the abundance for the specific line was deviating by more than 3$\sigma$ from the mean.

\section{Optimizing model atmosphere parameters}     \label{modelparam}

    As already mentioned in the introduction we started our abundance analysis with fundamental parameters estimated from Str\"{o}mgren photometry using the calibration of \citet{1985MNRAS.217..305M}, see Table\,\ref{photometryparams} for details. Furthermore, we took \vmic\,=\,2\,\kms and Z\,=\,0.2 as a first approximation.

    Microturbulence is usually determined by removing a trend of line abundances with equivalent widths. Due to the rather high rotational velocity of \HD\ a statistically significant sample of lines for measuring equivalent widths does not exist. We therefore replaced equivalent widths by the central line depths of unsaturated lines. To test this approach we determined equivalent widths and central line depths of nearly unblended synthetic line profiles (Z\,=\,0.2) using different microturbulent velocities from 0 to 4\,\kms\ and in 0.5\,\kms\ steps. All measurements correlated perfectly
to except for a constant off-set depending on the chosen microturbulence. We concluded that replacing the equivalent width by the central line depth is justified in our case. For Figure\,\ref{LdVsEqw} in the appendix we arbitrarily normalized the measurements to \vmic\,=\,0\,\kms.

    Element abundances for all 112 Fe\,I and Fe\,II lines were calculated and plotted versus their individual central line depth. The slope of the linear fit changes significantly with the microturbulence value used for the line-abundance determination, where a slope of zero indicates an optimum  microturbulent velocity. As a second indicator, the line-abundance scatter was used and plotted against \vmic. With the average value from both methods, a new atmosphere model was computed.

    This model was used to determine a first estimate of \logg\ from our spectra, with the other parameters unchanged. A consistent value for the surface gravity is found, when the line abundances for a given element are independent of the ionization stage. We used the resulting \logg-value for the next step, an improved estimate - beyond photometry - of the effective temperature.

    Lines with different excitation potential should give the same abundance, if a correct effective temperature was chosen for the model. Hence, this best fitting temperature value was found by eliminating any trend in the relation abundance-versus-excitation potential in our iron line sample.

The so determined set of fundamental parameters was then used in the following iterations according to the scheme: \vmic\ - \logg\ - \teff\ - \logg\ - \vmic\ - \logg\ - \teff\ - etc., until no significant changes occurred between subsequent iteration steps. During the process, we used ODF-model atmospheres with Kurucz ODF-tables (see section~\ref{Tools}).

After having determined an optimum ODF-based model atmosphere via iron lines we determined abundances of all other elements - where possible. The following procedure was used for selecting usable spectral lines: \\
$\bullet$\hspace{1mm} All lines between 4000\,{\AA}  and 7500\,{\AA} with a central line depth above a certain limit were selected. For elements with many usable  lines, 0.1 was the limit. For elements with fewer lines, 0.025 or even 0.01 was used and the wavelength range was occasionally extended to the entire observed spectrum. These are subjective values based on experience.\\
$\bullet$\hspace{1mm} The semi-automatic procedure \textit{linfit} from the ATC package was used to fit these lines, taking line blends into account. \\
 $\bullet$\hspace{1mm} Lines with an RMS error of $>1$ for the difference of observed and synthetic spectra were rejected. \\
$\bullet$\hspace{1mm} The standard deviation of the remaining sample of line abundances was calculated and all lines deviating by more than 3$\sigma$ were rejected. \\
$\bullet$\hspace{1mm}  A histogram with an appropriate abundance bin width was created. This width was typically 0.1 to 0.4\,dex per bin, depending on the number of usable lines.  \\
$\bullet$\hspace{1mm}  A Gaussian bell function was fitted to the histogram. The centre and FWHM to this fit were taken as average abundance and its error.

    \begin{table}
    \caption{Microturbulent velocities, surface gravities and effective temperatures for different elements, based on ODF model atmospheres. For Ti and Ni, different numbers of lines (N) were used for the determination of \vmic and \teff. The effective temperature based on Cr, as well as the gravity value from Ti were rejected as outliers. }
    \label{params_fin}
    \centering
    \begin{tabular}{l c c c c}
    \hline\hline
    Element, $\sigma$ & \vmic & \logg & \teff & N \\
    & [\kms] & & [K] & \\
    \hline
    Fe & 2.4 & 4.0 & 7480 & 99 \\
    $\sigma$ & 0.2 & & 180 & \\
    Cr & 3.6 & 4.1 & (8180) & 130 \\
    $\sigma$ & 0.4 & & & \\
    Ti & 3.1 & (4.8) & 7590 & 78/73 \\
    $\sigma$ & 0.4 & & 230 & \\
    Ni & 3.2 & & 7620 & 117/111 \\
    $\sigma$ & 0.3 & & 110 & \\
    \hline
    Average & 3.1 & 4.1 & 7560 & \\
    $\sigma$ & 0.5 & 0.2 & 180 & \\
    \hline
    \end{tabular}
    \end{table}

    In a next step we used these new abundances for spectral synthesis, still based on ODF-model atmospheres, and determined \vmic, \logg, and \teff\ once again, but separately for Cr, Ti and Ni. Averages and error estimates are listed in Tab.\,\ref{params_fin}.

\section{Abundance determination}   \label{abund}

      With the abundances determined so far and the new atmospheric parameters that are given in the bottom of Tab.\,\ref{params_fin}, a LL-stellar model was calculated, now with the use of individual elemental abundances and no longer solar scaled ODF tables. Once more, the abundances and atmospheric parameters for Fe, Cr, Ti and Ni were calculated.

Comparing the LL-model abundances for these elements to the abundances derived from the ODF-model with the same atmospheric parameters, the LL-model abundances are slightly lower by 0.1\,dex for all four elements (see Fig.\,\ref{odf_ll} in the appendix). This difference is small compared to the typical standard deviation of 0.3\,dex for line-abundances and consequently no further adjustments to the atmospheric parameters were necessary.

 The LL-model with the average atmospheric parameters from Tab.\,\ref{params_fin} was used for the abundance determination of all other elements. The line selection procedure, as described in the former section, was applied again with the exception that we used the standard deviation of line-abundances as abundance error instead of fitting Gaussian bell functions to a histogram. If the abundance of an element could not be determined by us due to a lack of usable spectral lines, the abundance according to Z\,=\,+0.12 was used for computing our LL-model atmosphere.

\section{Magnetic field}    \label{magfield}

\begin{figure}[h]
    \centering
    \includegraphics[width=7.5cm]{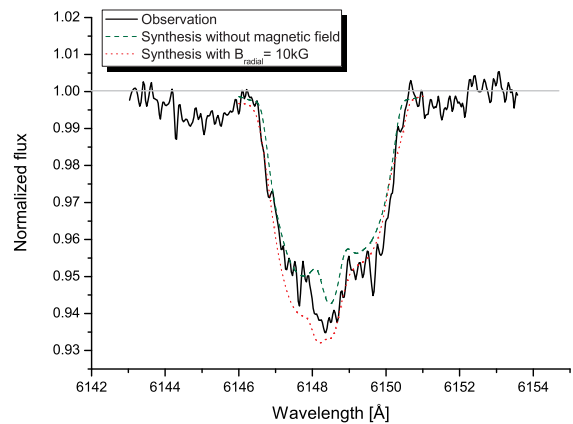}
          \caption{Observed spectrum around Fe\,II at 6149.258\,{\AA} including syntheses with and without magnetic field.}
          \label{mag_linesplit}
    \end{figure}

    We applied the spectrum synthesis code \textit{synth\_mag} to determine the magnetic field strength of \HD, using the magnetic field sensitive Fe\,II line at 6149.258\,{\AA} \citep{2003A&A...402..729S}, but obtained an inconclusive result (see Figure~\ref{mag_linesplit}).

    If a magnetic field is present, lines with a large Land\'{e} factor should provide a higher abundance for an analysis described in the present paper than those with a small factor. We calculated linear fits to the line-abundance versus Land\'e factor relation individually for Fe, Cr, Ti, Ni, Ca, Co, and Nd and obtained an average slope of -0.016 with an average error of 0.102 and a correlation coefficient R$^2\,=\,0.019$.  The low value for R$^2$ and the (small) negative slope indicate a magnetic field not measurable with our data. Above all, a magnetic field should create a positive slope.

\begin{figure}
    \centering
    \includegraphics[width=7.5cm]{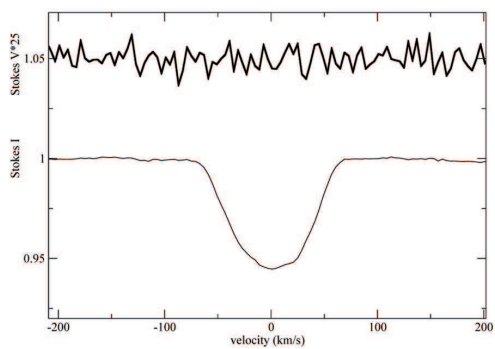}
          \caption{LSD profiles for Stokes I and V. The profile for Stokes V was multiplied by a factor of 25 for better visibility. No polarimetric signal is evident, which indicates the absence of a magnetic field exceeding 100\,Gauss.}
          \label{lsd}
    \end{figure}

    In order to completely rule out a measurable magnetic field, polarimetric observations were carried out by M. Gruberbauer at the OHP and analyzed by G.A. Wade. They used the LSD technique and a line mask for an 8000\,K star. The result is consistent with a null result for a longitudinal magnetic field. Figure~\ref{lsd} shows the LSD profiles for Stokes I and V.

%__________________________________________________________________
\section{Abundance error discussion}      \label{error}
When performing abundance analyses, it is important to keep control over the effects of all possible sources of error on the results. We therefore estimated their influence on the final abundances by changing parameters individually and by looking for the resulting deviations in the abundance determination.

\subsection{Continuum normalization}
    The continuum normalization is one of the main error sources in an abundance analysis.
    To investigate this effect the observed continuum normalized spectrum was off-set by 1\% to smaller and larger relative intensities. For the iron lines of Tab.\,\ref{vrad_individual} abundances were calculated for the shifted observations and compared to the unshifted results.

    \begin{figure}[h]
    \centering
    \includegraphics[width=7.5cm]{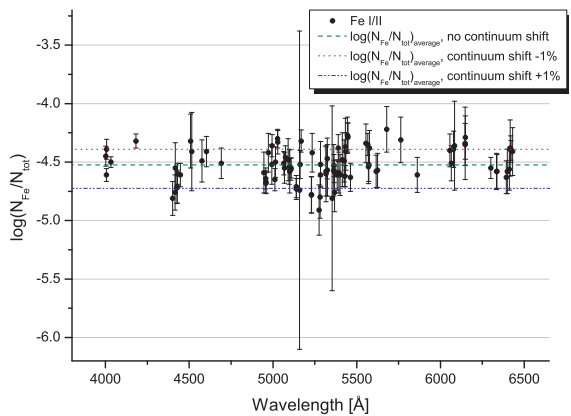}
          \caption{Iron line-abundances with errors due to a continuum normalization off-set by $\pm$1\%.}
          \label{err_norm}
    \end{figure}

Figure~\ref{err_norm} shows the iron line-abundances with error bars due to the continuum shifts. Averages are marked by the dashed, dotted and dashed-dotted horizontal lines.
    Some lines show an extraordinarily large error which is caused by the inability of the algorithm to fit the specific line when a shift to the continuum is applied. The typical error is 0.17\,dex with a minimum at 0.05\,dex and a maximum at 1.36\,dex, with the latter being much larger than the error of the entire line sample (0.15\,dex).

As the real continuum error varies throughout the spectrum in positive and negative directions, no change in the average abundance is expected in our case for a reasonably small error in the continuum determination. Only the scatter of the line-abundances will be increased relative to the case of a better normalized spectrum.

\subsection{Continuum S/N ratio}
    Noisy observations also increase errors in the abundance determination. To quantify this error source we computed rotationally broadened synthetic spectra from 4000\,\AA\ -- 4100\,{\AA} and added normally distributed noise, which resulted in 7 synthetic spectra with different S/N ratios in the continuum. Iron lines with a central line depth larger than 0.3 (without rotational broadening!) were selected, automatically fitted and the results are summarized in Tab.\,\ref{err_snr}.

    \begin{table}[h]
    \caption{Iron abundances determined from noisy synthetic spectra with atmospheric parameters according to \HD. Cont.-S/N: signal-to-noise ratio of the continuum.}
    \label{err_snr}
    \centering
    \begin{tabular}{c c c}
    \hline\hline
    Cont.-S/N & log(N$_{Fe}$/N$_{tot}$) & Error ($\sigma$) \\
    \hline
    10 & -4.526 & 0.189 \\
    25 & -4.520 & 0.087 \\
    50 & -4.520 & 0.035 \\
    100 & -4.518 & 0.010 \\
    200 & -4.519 & 0.007 \\
    300 & -4.519 & 0.004 \\
    $\infty$ & -4.520 & 0.000 \\
    \hline
    \end{tabular}
    \end{table}

    The average S/N ratio of the observations for \HD\ can be estimated to be above 150 for the mentioned region, which corresponds to an error in abundance for the stronger lines of less than 0.01 dex. For the weaker lines, this error obviously will increase. Lines with a central depth of only 0.1 (again without rotational broadening!) have errors of up to 0.05\,dex for a S/N ratio of 100.

\subsection{Projected rotational velocity}
    An error in the projected rotational velocity affects the abundance determination, because the line fitting algorithm compensates an incorrect line depth with a change of abundance. A lower value of \vsin\ produces deeper, a higher value shallower synthetic line cores. Therefore, the line (core) fitting algorithm would calculate a wrong abundance value for the line. This effect causes a systematical over- (in case of an overestimation of \vsin\,), or underestimation (in case of an underestimation of \vsin\,) of all element abundances. To estimate the size of this effect, the mean iron abundance based on the sample of 99 lines was calculated for two models with a \vsin\ of 48.1 and 58.1\,\kms. We then applied a linear fit to the plot element abundance vs. projected rotational velocity. This allowed us to transform the uncertainty in \vsin\ of 0.5\,\kms\ to an Fe--abundance error of 0.01\,dex, which is also a typical value for the other metals.

\subsection{Model atmosphere parameters}
    The abundance errors within the line-samples cause errors in the calculated slopes of the abundance vs. parameter fits.

\begin{figure}
    \centering
    \includegraphics[width=7.5cm]{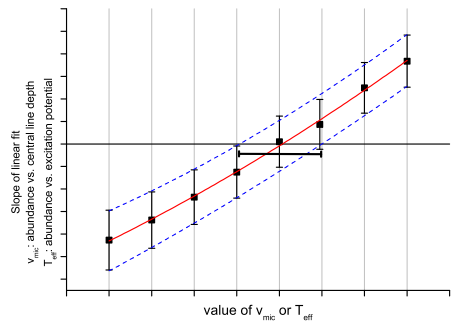}
          \caption{Error estimation principle demonstrated for \vmic\ and \teff. The dots give the computed slope of a linear fit to line-abundances versus central line depth (in the case of \vmic\ determination), respectively, line-abundance versus excitation potential (in the case of \teff\ determination). The vertical error bars represent the slope-errors for the respective linear fits. The horizontal error bar close to the x-axis denotes the adopted error in the fundamental parameter.}
          \label{error_estimation}
    \end{figure}

    Polynomials (maximum order 2) were fitted through the upper and lower limits of these slope errors and their intersection with the x-axis was computed. Half of the distance between the two intersections is used by us as a reasonable error estimate for the determined parameter. The principle is illustrated in Figure~\ref{error_estimation}.

\begin{figure}
    \centering
    \includegraphics[width=7.5cm]{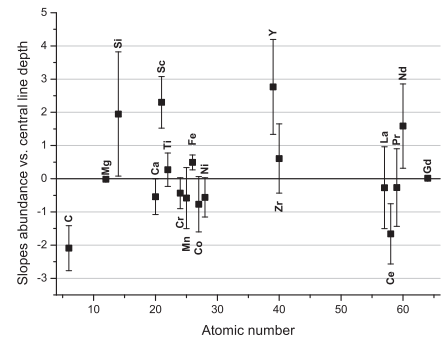}
          \caption{Slopes for different elements of linear fits of abundance vs. central line depth for \vmic\,=\,3.1\,\kms.}
          \label{err_abuvmic}
    \end{figure}

\begin{figure}
    \centering
    \includegraphics[width=7.5cm]{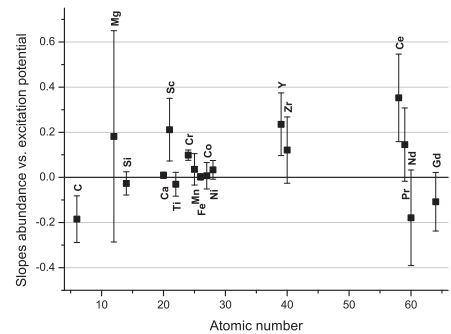}
          \caption{Slopes for different elements of linear fits of abundance vs. excitation potential for \teff\,=\,7560\,K.}
          \label{err_abuteff}
    \end{figure}

\begin{figure}
    \centering
    \includegraphics[width=7.5cm]{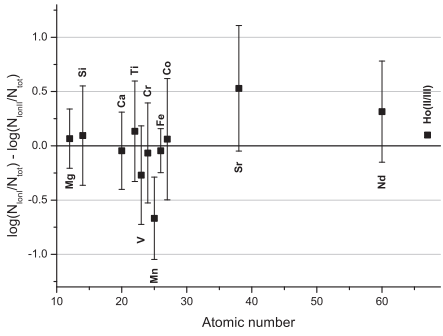}
          \caption{Abundance differences (neutral minus singly ionized species) for \logg\,=\,4.1 .}
          \label{err_abulogg}
    \end{figure}

    Based on two independently determined values for \logg\ of 4.0 for Fe and 4.1 for Cr, both confirming the value of 4.1 from photometry, we adopted a conservative error of 0.2 for \logg.
Experience shows, that for a typical standard deviation in abundance within the iron line sample of 0.15\,dex in a faster rotating star, this value is reasonable.
    For elements with a sufficiently large number of lines, slopes of abundance-versus-central line depth (\vmic\,-\,sensitive, Figure~\ref{err_abuvmic}), abundance-versus-excitation potential (\teff\,-\,sensitive, Figure~\ref{err_abuteff}) and abundance differences between the ionization stages (\logg\,-\,sensitive, Figure~\ref{err_abulogg}) were calculated in order to check the adopted final fundamental parameters, which were based mainly on Fe, Cr, Ti, and Ni lines, as is described in Sec.\,\ref{abund}.


\subsection{Error estimate for abundances}
    Similar to the error estimate for the final atmospheric parameters, we investigated the influence of microturbulence, gravity and effective temperature on the derived abundances.

    For this purpose, the fundamental parameters of the synthetic spectrum were changed by $\pm$ the estimated parameter error -- as was determined in the previous subsection -- and then the synthetic line profiles fitted to the observations by changing the abundances.
    Applying a linear fit to the plots abundance-versus-parameter value gives an error estimate for the element abundances. Table~\ref{err_abus} shows for iron the resulting error budget, which is typical for the other elements.

    \begin{table}[h]
    \caption{Influence of the individual parameter errors on the abundance determination. This particular Table was produced for iron, but is representative for the other elements.}
    \label{err_abus}
    \centering
    \begin{tabular}{c c c c}
    \hline\hline
    Parameter & Value & Error & Fe$_{error}$ [dex] \\
    \hline
    \teff\,[K] & 7560 & 180 & 0.07 \\
    \logg\, & 4.1 & 0.2 & 0.005 \\
    \vmic\,[\kms] & 3.1 & 0.5 & 0.01 \\
    \hline
    \end{tabular}
    \end{table}

\section{Results and Discussion}            \label{Results}

\begin{figure}
    \centering
    \includegraphics[width=7.5cm]{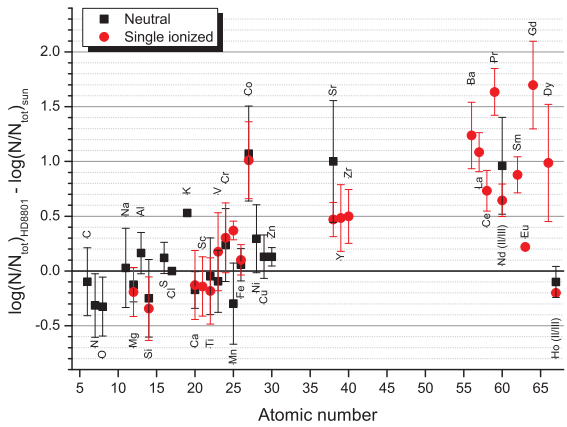}
          \caption{Abundance differences relative to the Sun for individual ionization stages. The solar abundances are adopted from \citet{2005ASPC..336...25A} and \citet{2001AIPC..598...23H}.}
          \label{abus_ions_fig}
    \end{figure}

Table~\ref{abus_all} shows the final results of our abundance analysis and Fig.\,\ref{abus_ions_fig} is a graphical representation of the abundances relative to the Sun. The analysis was based on the following fundamental parameters for \HD: \teff\,=\,7560\,K, \logg\,=\,4.1, \vmic\,=\,3.1\,\kms, which were derived from spectroscopy, and a model atmosphere computed with the element abundances determined by us.

    \begin{table}[h]
    \caption{Average abundances with standard deviations and number of used lines (N) for neutral(I) and single ionized(II) elements. For Nd and Ho (*) the ionization stages are II and III. Abundance values in columns I and II are (N/N$_{tot}$) in logarithmic scale. Column Sun contains the solar abundances for Fig.\,\ref{abus_ions_fig} and Fig.\,\ref{abuam} taken from \citet{2005ASPC..336...25A} and \citet{2001AIPC..598...23H}.}
    \label{abus_all}
    \centering
    \begin{tabular}{l c c c c c c c}
    \hline\hline
    El. & I & $\sigma$ & N & II & $\sigma$ & N & Sun \\
    \hline
    C & -3.75 & 0.31 & 42 &&&& -3.65 \\
    N & -4.58 & 0.29 & 4 &&&& -4.26 \\
    O & -3.71 & 0.27 & 10 &&&& -3.38 \\
    Na & -5.84 & 0.36 & 5 &&&& -5.87 \\
    Mg & -4.66 & 0.18 & 9 & -4.70 & 0.22 & 6 & -4.51 \\
    Al & -5.51 & 0.19 & 3 &&&& -5.67 \\
    Si & -4.78 & 0.35 & 18 & -4.87 & 0.29 & 3 & -4.53 \\
    S & -4.78 & 0.14 & 10 &&&& -4.9 \\
    Cl & -6.54 &  & 1 &&&& -6.54 \\
    K & -6.43 &  & 1 &&&& -6.96 \\
    Ca & -5.90 & 0.17 & 35 & -5.86 & 0.31 & 9 & -5.73 \\
    Sc & &&& -9.13 & 0.27 & 17 & -8.99 \\
    Ti & -7.19 & 0.35 & 18 & -7.32 & 0.30 & 57 & -7.14 \\
    V & -8.13 & 0.28 & 5 & -7.86 & 0.36 & 10 & -8.04 \\
    Cr & -6.16 & 0.33 & 76 & -6.10 & 0.32 & 46 & -6.4 \\
    Mn & -6.95 & 0.37 & 28 & -6.28 & 0.08 & 2 & -6.65 \\
    Fe & -4.53 & 0.15 & 79 & -4.49 & 0.14 & 20 & -4.59 \\
    Co & -6.05 & 0.43 & 69 & -6.11 & 0.35 & 10 & -7.12 \\
    Ni & -5.51 & 0.31 & 110 &&&& -5.81 \\
    Cu & -7.70 & 0.20 & 2 &&&& -7.83 \\
    Zn & -7.31 & 0.08 & 2 &&&& -7.44 \\
    Sr & -8.12 & 0.56 & 4 & -8.65 & 0.16 & 2 & -9.12 \\
    Y & &&& -9.35 & 0.30 & 10 & -9.83 \\
    Zr & &&& -8.95 & 0.25 & 14 & -9.45 \\
    Ba & &&& -8.63 & 0.30 & 4 & -9.87 \\
    La & &&& -9.83 & 0.18 & 9 & -10.91 \\
    Ce & &&& -9.73 & 0.19 & 16 & -10.46 \\
    Pr & &&& -9.70 & 0.21 & 13 & -11.33 \\
    Nd* & -9.63 & 0.44 & 31 & -9.95 & 0.15 & 2 & -10.59 \\
    Sm & &&& -10.15 & 0.17 & 5 & -11.03 \\
    Eu & &&& -11.30 &  & 1 & -11.52 \\
    Gd & &&& -9.22 & 0.40 & 37 & -10.92 \\
    Dy & &&& -9.91 & 0.54 & 9 & -10.9 \\
    Ho* & -11.66 & 0.14 & 2 & -11.73 & & 1 & -11.53 \\
    \hline
    \end{tabular}
    \end{table}

\begin{figure}[h]
    \centering
    \includegraphics[width=7.5cm]{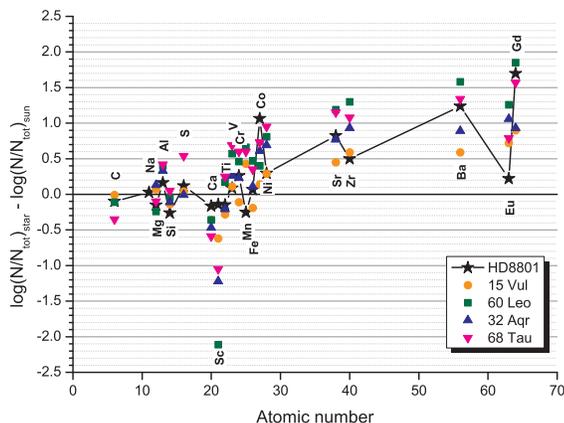}
          \caption{Abundance pattern for different Am stars. The stars denote the abundances derived for \HD\ in this analysis.}
          \label{abuam}
    \end{figure}

    \HD\ shows an abundance pattern, very similar to Am stars except of the higher abundance of scandium. A comparison with other Am stars is shown in Figure~\ref{abuam}. The abundances for the 4 other stars were taken from \citet{1997MNRAS.288..470A}, \citet{1999MNRAS.305..591A} and \citet{1992MNRAS.258..270B}. We attempted to determine the integrated stellar surface magnetic field, but arrived only at an upper limit of less than 100\, Gauss, which is in agreement with Am stars having no measurable magnetic field.

\HD\ is remarkable in several respects. It does not only show $\gamma$ Doradus and $\delta$ Scuti pulsation, but also shows oscillations with periods in between the two clearly separated domains, whose excitation requires explanation. Marginal and evolved Am stars are known to pulsate, whereas the pulsational incidence in unevolved classical Am stars is quite low (e.g., \citet{1989MNRAS.238.1077K}). Yet our analysis clearly showed that the hybrid pulsator \HD\ is a main-sequence Am star. Finally, about 75\% of Am stars are located in binaries with orbital periods shorter than three years \citep{1985ApJS...59..229A}. There is no evidence whatsoever that \HD\ has a binary companion.

Comparing \HD\ to other hybrid pulsators with detailed abundance analyses, HD 114839 is a chemically normal main sequence star that shows oscillations between the $\gamma$ Doradus and $\delta$ Scuti domains, whereas BD+18 4914 does not show such oscillations, but is a mildly evolved Am star \citep{2011ApJ...743..153H}. The behaviour of only these three stars is already sufficiently contradictory to conclude that it is unlikely that there will be a clear connection between Am spectral type and pulsational excitation, at least as far as hybrid pulsators are concerned. We do however note that \citet{2000A&A...360..603T} suggested that diffusion, that is believed to be responsible for Am-type abundance anomalies may support the excitation of g-mode pulsations. A larger fraction of Am stars among $\gamma$ Doradus stars than among $\delta$ Scuti will be a hint in this direction. Finally, we remark that we do not consider two other hybrids here: due to the fast rotation of HD 49434 \citep{2008A&A...489.1213U} it cannot be safely said that it is a hybrid pulsator, whereas for the hybrid pulsator HD 178327 \citep{2011MNRAS.414..792B} the Am classification has been disputed \citep{2011ApJ...743..153H}. \HD\ therefore remains a unique object, and its rich pulsation spectrum in combination with the chemical peculiarity may justify a large observational effort to sound its interior.

\begin{acknowledgements}
    We wish to thank M. Gruberbauer and G. Wade for providing the spectropolarimetry and C. St\"{u}tz for his support during the analysis. This work was supported by the Austrian Science Fund (grants P22691-N16 and R12-N02).
\end{acknowledgements}

\bibliographystyle{aa} % style aa.bst
\bibliography{hd8801} % hd8801.bib

\begin{thebibliography}{38}
\expandafter\ifx\csname natexlab\endcsname\relax\def\natexlab#1{#1}\fi

\bibitem[{{Abt} \& {Levy}(1985)}]{1985ApJS...59..229A}
{Abt}, H.~A. \& {Levy}, S.~G. 1985, \apjs, 59, 229

\bibitem[{{Adelman} {et~al.}(1999){Adelman}, {Caliskan}, {Cay}, {Kocer}, \&
  {Tektanali}}]{1999MNRAS.305..591A}
{Adelman}, S.~J., {Caliskan}, H., {Cay}, T., {Kocer}, D., \& {Tektanali}, H.~G.
  1999, \mnras, 305, 591

\bibitem[{{Adelman} {et~al.}(1997){Adelman}, {Caliskan}, {Kocer}, \&
  {Bolcal}}]{1997MNRAS.288..470A}
{Adelman}, S.~J., {Caliskan}, H., {Kocer}, D., \& {Bolcal}, C. 1997, \mnras,
  288, 470

\bibitem[{{Asplund} {et~al.}(2005){Asplund}, {Grevesse}, \&
  {Sauval}}]{2005ASPC..336...25A}
{Asplund}, M., {Grevesse}, N., \& {Sauval}, A.~J. 2005, in Astronomical Society
  of the Pacific Conference Series, Vol. 336, Cosmic Abundances as Records of
  Stellar Evolution and Nucleosynthesis, ed. {T.~G.~Barnes III \& F.~N.~Bash},
  25--+

\bibitem[{{Balona}(1994)}]{1994MNRAS.268..119B}
{Balona}, L.~A. 1994, \mnras, 268, 119

\bibitem[{{Balona} {et~al.}(2011){Balona}, {Ripepi}, {Catanzaro}, {Kurtz},
  {Smalley}, {De Cat}, {Eyer}, {Grigahc{\`e}ne}, {Leccia}, {Southworth},
  {Uytterhoeven}, {van Winckel}, {Christensen-Dalsgaard}, {Kjeldsen},
  {Caldwell}, {van Cleve}, \& {Girouard}}]{2011MNRAS.414..792B}
{Balona}, L.~A., {Ripepi}, V., {Catanzaro}, G., {et~al.} 2011, \mnras, 414, 792

\bibitem[{{Bolcal} {et~al.}(1992){Bolcal}, {Kocer}, \&
  {Adelman}}]{1992MNRAS.258..270B}
{Bolcal}, C., {Kocer}, D., \& {Adelman}, S.~J. 1992, \mnras, 258, 270

\bibitem[{{Canuto} \& {Mazzitelli}(1991)}]{1991ApJ...370..295C}
{Canuto}, V.~M. \& {Mazzitelli}, I. 1991, \apj, 370, 295

\bibitem[{{Castelli} {et~al.}(1997){Castelli}, {Gratton}, \&
  {Kurucz}}]{1997A&A...318..841C}
{Castelli}, F., {Gratton}, R.~G., \& {Kurucz}, R.~L. 1997, \aap, 318, 841

\bibitem[{{Chapellier} {et~al.}(2012){Chapellier}, {Mathias}, {Weiss}, {Le
  Contel}, \& {Debosscher}}]{2012A&A...540A.117C}
{Chapellier}, E., {Mathias}, P., {Weiss}, W.~W., {Le Contel}, D., \&
  {Debosscher}, J. 2012, \aap, 540, A117

\bibitem[{{Fresno}(1994)}]{1994}
{Fresno}, E.~M. 1994, Universidad de Barcelona

\bibitem[{{Handler}(2009)}]{2009MNRAS.398.1339H}
{Handler}, G. 2009, \mnras, 398, 1339

\bibitem[{{Hareter} {et~al.}(2011){Hareter}, {Fossati}, {Weiss}, {Su{\'a}rez},
  {Uytterhoeven}, {Rainer}, \& {Poretti}}]{2011ApJ...743..153H}
{Hareter}, M., {Fossati}, L., {Weiss}, W., {et~al.} 2011, \apj, 743, 153

\bibitem[{{Heiter} {et~al.}(2008){Heiter}, {Barklem}, {Fossati}, {Kildiyarova},
  {Kochukhov}, {Kupka}, {Obbrugger}, {Piskunov}, {Plez}, {Ryabchikova},
  {Stempels}, {St{\"u}tz}, \& {Weiss}}]{2008JPhCS.130a2011H}
{Heiter}, U., {Barklem}, P., {Fossati}, L., {et~al.} 2008, Journal of Physics
  Conference Series, 130, 012011

\bibitem[{{Henry} \& {Fekel}(2005)}]{2005AJ....129.2026H}
{Henry}, G.~W. \& {Fekel}, F.~C. 2005, \aj, 129, 2026

\bibitem[{{Henry} {et~al.}(2011){Henry}, {Fekel}, \&
  {Henry}}]{2011AJ....142...39H}
{Henry}, G.~W., {Fekel}, F.~C., \& {Henry}, S.~M. 2011, \aj, 142, 39

\bibitem[{{Holweger}(2001)}]{2001AIPC..598...23H}
{Holweger}, H. 2001, in American Institute of Physics Conference Series, Vol.
  598, Joint SOHO/ACE workshop ''Solar and Galactic Composition'', ed.
  {R.~F.~Wimmer-Schweingruber}, 23--30

\bibitem[{{Johnson} \& {Morgan}(1953)}]{1953ApJ...117..313J}
{Johnson}, H.~L. \& {Morgan}, W.~W. 1953, \apj, 117, 313

\bibitem[{{Kaiser}(2006)}]{2006ASPC..349..257K}
{Kaiser}, A. 2006, in Astronomical Society of the Pacific Conference Series,
  Vol. 349, Astrophysics of Variable Stars, ed. {C.~Aerts \& C.~Sterken},
  257--+

\bibitem[{{Kochukhov}(2007)}]{2007pms..conf..109K}
{Kochukhov}, O.~P. 2007, in Physics of Magnetic Stars, ed. {I.~I.~Romanyuk,
  D.~O.~Kudryavtsev, O.~M.~Neizvestnaya, \& V.~M.~Shapoval }, 109--118

\bibitem[{{Kupka} {et~al.}(1999){Kupka}, {Piskunov}, {Ryabchikova}, {Stempels},
  \& {Weiss}}]{1999A&AS..138..119K}
{Kupka}, F., {Piskunov}, N., {Ryabchikova}, T.~A., {Stempels}, H.~C., \&
  {Weiss}, W.~W. 1999, \aaps, 138, 119

\bibitem[{{Kupka} {et~al.}(2000){Kupka}, {Ryabchikova}, {Piskunov}, {Stempels},
  \& {Weiss}}]{2000BaltA...9..590K}
{Kupka}, F.~G., {Ryabchikova}, T.~A., {Piskunov}, N.~E., {Stempels}, H.~C., \&
  {Weiss}, W.~W. 2000, Baltic Astronomy, 9, 590

\bibitem[{{Kurtz}(1998)}]{1998ASPC..135..420K}
{Kurtz}, D. 1998, in Astronomical Society of the Pacific Conference Series,
  Vol. 135, A Half Century of Stellar Pulsation Interpretation, ed. P.~A.
  {Bradley} \& J.~A. {Guzik}, 420

\bibitem[{{Kurtz}(1989)}]{1989MNRAS.238.1077K}
{Kurtz}, D.~W. 1989, \mnras, 238, 1077

\bibitem[{{Moon} \& {Dworetsky}(1985)}]{1985MNRAS.217..305M}
{Moon}, T.~T. \& {Dworetsky}, M.~M. 1985, \mnras, 217, 305

\bibitem[{{Napiwotzki} {et~al.}(1993){Napiwotzki}, {Schoenberner}, \&
  {Wenske}}]{1993A&A...268..653N}
{Napiwotzki}, R., {Schoenberner}, D., \& {Wenske}, V. 1993, \aap, 268, 653

\bibitem[{{Nesvacil} {et~al.}(2003){Nesvacil}, {St{\"u}tz}, \&
  {Weiss}}]{2003ASPC..298..173N}
{Nesvacil}, N., {St{\"u}tz}, C., \& {Weiss}, W.~W. 2003, in Astronomical
  Society of the Pacific Conference Series, Vol. 298, GAIA Spectroscopy:
  Science and Technology, ed. {U.~Munari}, 173--+

\bibitem[{{Olsen}(1984)}]{1984A&AS...57..443O}
{Olsen}, E.~H. 1984, \aaps, 57, 443

\bibitem[{{Piskunov}(1992)}]{1992pess.conf...92P}
{Piskunov}, N.~E. 1992, in Physics and Evolution of Stars: Stellar Magnetism,
  ed. {Y.~V.~Glagolevskij \& I.~I.~Romanyuk}, 92

\bibitem[{{Piskunov} {et~al.}(1995){Piskunov}, {Kupka}, {Ryabchikova}, {Weiss},
  \& {Jeffery}}]{1995A&AS..112..525P}
{Piskunov}, N.~E., {Kupka}, F., {Ryabchikova}, T.~A., {Weiss}, W.~W., \&
  {Jeffery}, C.~S. 1995, \aaps, 112, 525

\bibitem[{{Ribas} {et~al.}(1997){Ribas}, {Jordi}, {Torra}, \&
  {Gimenez}}]{1997A&A...327..207R}
{Ribas}, I., {Jordi}, C., {Torra}, J., \& {Gimenez}, A. 1997, \aap, 327, 207

\bibitem[{{Ryabchikova} {et~al.}(1997){Ryabchikova}, {Piskunov}, {Kupka}, \&
  {Weiss}}]{1997BaltA...6..244R}
{Ryabchikova}, T.~A., {Piskunov}, N.~E., {Kupka}, F., \& {Weiss}, W.~W. 1997,
  Baltic Astronomy, 6, 244

\bibitem[{{Schuster} \& {Nissen}(1989)}]{1989A&A...221...65S}
{Schuster}, W.~J. \& {Nissen}, P.~E. 1989, \aap, 221, 65

\bibitem[{{Shulyak} {et~al.}(2004){Shulyak}, {Tsymbal}, {Ryabchikova},
  {St{\"u}tz}, \& {Weiss}}]{2004A&A...428..993S}
{Shulyak}, D., {Tsymbal}, V., {Ryabchikova}, T., {St{\"u}tz}, C., \& {Weiss},
  W.~W. 2004, \aap, 428, 993

\bibitem[{{St{\"u}tz}(2005)}]{2005MSAIS...8..165S}
{St{\"u}tz}, C. 2005, Memorie della Societa Astronomica Italiana Supplement, 8,
  165

\bibitem[{{St{\"u}tz} {et~al.}(2003){St{\"u}tz}, {Ryabchikova}, \&
  {Weiss}}]{2003A&A...402..729S}
{St{\"u}tz}, C., {Ryabchikova}, T., \& {Weiss}, W.~W. 2003, \aap, 402, 729

\bibitem[{{Turcotte} {et~al.}(2000){Turcotte}, {Richer}, {Michaud}, \&
  {Christensen-Dalsgaard}}]{2000A&A...360..603T}
{Turcotte}, S., {Richer}, J., {Michaud}, G., \& {Christensen-Dalsgaard}, J.
  2000, \aap, 360, 603

\bibitem[{{Uytterhoeven} {et~al.}(2008){Uytterhoeven}, {Mathias}, {Poretti},
  {Rainer}, {Mart{\'{\i}}n-Ruiz}, {Rodr{\'{\i}}guez}, {Amado}, {Le Contel},
  {Jankov}, {Niemczura}, {Pollard}, {Brunsden}, {Papar{\'o}}, {Costa},
  {Valtier}, {Garrido}, {Su{\'a}rez}, {Kilmartin}, {Chapellier},
  {Rodr{\'{\i}}guez-L{\'o}pez}, {Marin}, {Aceituno}, {Casanova}, {Rolland}, \&
  {Olivares}}]{2008A&A...489.1213U}
{Uytterhoeven}, K., {Mathias}, P., {Poretti}, E., {et~al.} 2008, \aap, 489,
  1213

\end{thebibliography}
\newpage

\appendix

\section{Additional tables and figures}
     \begin{table}[h]
    \caption{Radial velocities derived for different Echelle orders by cross correlating with a synthetic spectrum.
    R$_{max}$ shows the maximum cross correlation coefficient (see section~\ref{Radial velocity}), numLin shows the number of lines used.}
    \label{vrad}
    \centering
    \begin{tabular}{c c c c c c}
    \hline\hline
    order & from & to & \vrad & R$_{max}$ & numLin  \\
    & {[{\AA}]} & [{\AA}] & [\kms] & & (see Tab.\,\ref{vrad_individual}) \\
    \hline
    1 & 4000 & 4047 & -22.037 & 0.8044 & 1 \\
    2 & 4050 & 4093 & -19.567 & 0.7799 & 1 \\
    3 & 4148 & 4196 & -24.994 & 0.7632 & 2 \\
    4 & 4197 & 4246 & -24.163 & 0.8381 & 1 \\
    5 & 4250 & 4310 & -21.743 & 0.8424 & 1 \\
    6 & 4367 & 4405 & -23.485 & 0.8614 & 1 \\
    7 & 4423 & 4464 & -23.448 & 0.8728 & 1 \\
    8 & 4466 & 4522 & -24.535 & 0.9202 & 4 \\
    9 & 4524 & 4581 & -22.397 & 0.8209 & 4 \\
    10 & 4594 & 4694 & -26.779 & 0.7704 & 1 \\
    11 & 4930 & 5043 & -21.124 & 0.9174 & 1 \\
    12 & 5046 & 5199 & -22.406 & 0.9441 & 3 \\
    13 & 5212 & 5281 & -24.242 & 0.9601 & 2 \\
    14 & 5290 & 5358 & -23.343 & 0.9229 & 3 \\
    15 & 5375 & 5435 & -21.912 & 0.9561 & 3 \\
    16 & 5460 & 5544 & -21.660 & 0.7455 & 2 \\
    17 & 5547 & 5597 & -22.182 & 0.7776 & 2 \\
    18 & 5638 & 5703 & -24.646 & 0.5488 & 0 \\
    19 & 5733 & 5821 & -22.819 & 0.2563 & 0 \\
    20 & 6034 & 6126 & -25.124 & 0.3572 & 2 \\
    21 & 6142 & 6234 & -22.360 & 0.4573 & 1 \\
    22 & 6373 & 6453 & -21.828 & 0.5016 & 3 \\
    23 & 6637 & 6713 & -24.004 & 0.1904 & 0 \\
    24 & 6744 & 6846 & -21.319 & 0.1346 & 0 \\
    25 & 7477 & 7590 & -23.809 & 0.2830 & 1 \\
    26 & 8391 & 8517 & -26.392 & 0.6112 & 0 \\
    \hline
    \end{tabular}
    \end{table}

   \begin{table}[h]
    \caption{Average radial velocities and errors using various weights derived by cross correlation with a synthetic spectrum, and from individual spectral lines listed in Tab.\,\ref{vrad_individual}.}
    \label{vrad_ave}
    \centering
    \begin{tabular}{l c c}
    \hline\hline
    & \vrad & $\sigma(weighted)$ \\
    & [\kms] & [\kms] \\
    \hline
    individual lines (Tab.\,\ref{vrad_individual})& -22.7 & 1.7 \\
    weight: none & -23.2 & 1.7 \\
    weight: R$_{max}$ & -23.1 & 1.7 \\
    weight: number of lines  & -23.1 & 1.5 \\
    \hline
    adopted value: & -23.1 & 1.6 \\
    \end{tabular}
    \end{table}

    \begin{figure}[h]
    \centering
    \includegraphics[width=7.5cm]{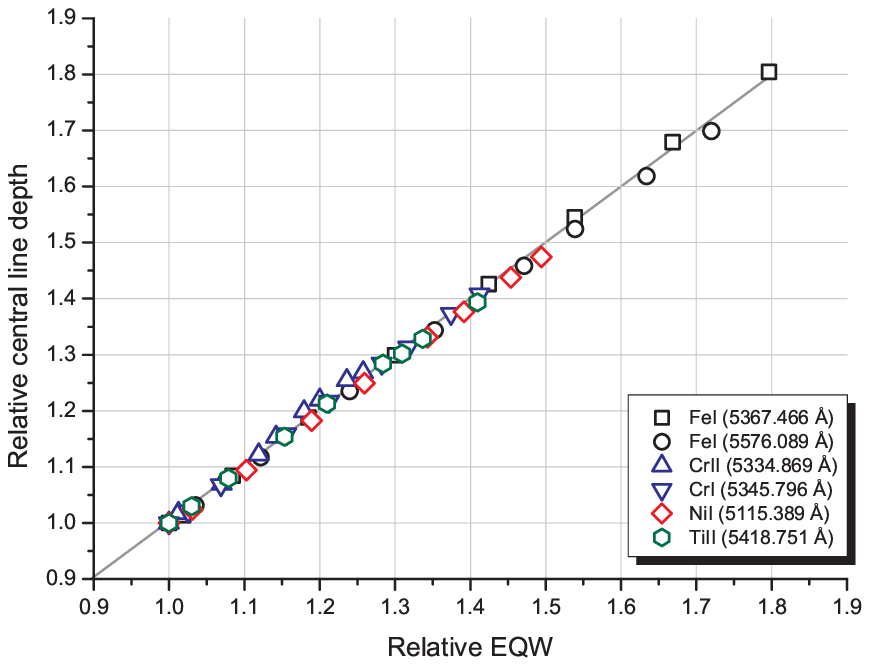}
          \caption{Relative central line depths versus equivalent widths for different unblended synthetic spectral lines.
          Microturbulent velocities from 0 to 4 \kms were used for this comparison, which resulted for different velocities in constant off-sets. For this Figure the relative central line depths were normalized to the case of \vmic\,=\,0\,\kms.}
          \label{LdVsEqw}
    \end{figure}

   \begin{figure}
    \centering
    \includegraphics[width=7.5cm]{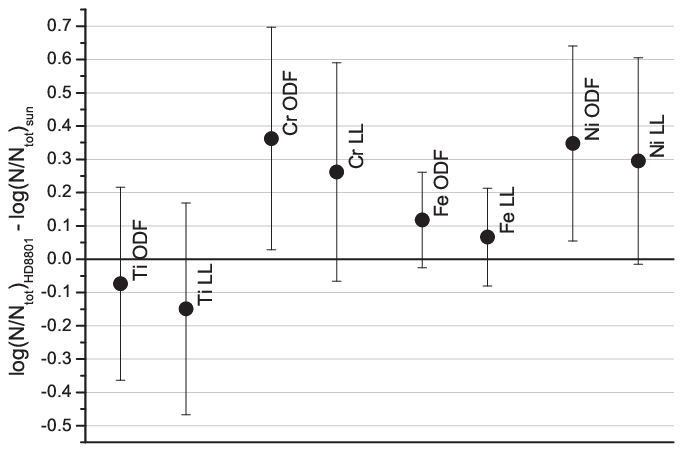}
          \caption{Abundances based on the final ODF-model with Z = +0.12 and the LL-model based on individual abundances as is described in Sec.\,\ref{abund} and listed in Tab.\,\ref{abus_all}.
          }
          \label{odf_ll}
    \end{figure}
\end{document}